\documentclass[runningheads]{llncs}

\usepackage[T1]{fontenc}
\usepackage[utf8]{inputenc}
\usepackage{acro}
\usepackage{adjustbox}
\usepackage{booktabs}
\usepackage{comment}
\usepackage{todonotes}
\usepackage{colortbl}

\newcolumntype{a}{>{\columncolor{gray!10!white}}c}
\newcolumntype{x}{>{\columncolor{green!10!white}}c}
\newcolumntype{y}{>{\columncolor{blue!10!white}}c}
\newcolumntype{z}{>{\columncolor{yellow!10!white}}c}
\newcolumntype{v}{>{\columncolor{red!10!white}}c}
\definecolor{OliveGreen}{rgb}{0,0.6,0}
\definecolor{ForestGreen}{RGB}{34,139,34}
\definecolor{myblue}{RGB}{37,165,203}
\definecolor{FAUblue}{rgb}{0.000, 0.2196, 0.3961}
\definecolor{myred}{RGB}{175,32,67}

\colorlet{backgroundcol}{cyan!10!white}

\usepackage{graphicx}
\usepackage{color}
\usepackage{xcolor}
\usepackage{listings}
\usepackage[hidelinks]{hyperref}

\definecolor{codegreen}{rgb}{0,0.6,0}
\definecolor{codegray}{rgb}{0.5,0.5,0.5}
\definecolor{codepurple}{rgb}{0.58,0.5,0.82}
\definecolor{backcolour}{rgb}{0.95,0.95,0.92}

\lstdefinestyle{mystyle}{
    backgroundcolor=\color{backcolour},   
    commentstyle=\color{codegreen},
    keywordstyle=\color{magenta}\bfseries,
    moredelim=[is][\color{magenta}\bfseries]{@}{@},
    numberstyle=\tiny\color{codegray},
    stringstyle=\color{codepurple},
    basicstyle=\C\footnotesize,
    breakatwhitespace=false,         
    breaklines=true,                 
    captionpos=b,                    
    keepspaces=true,                 
    numbers=left,                    
    numbersep=5pt,                  
    showspaces=false,                
    showstringspaces=false,
    showtabs=false,                  
    tabsize=2
}

\usepackage{amsmath,amssymb,amsfonts}

\usepackage{wrapfig}
\usepackage{graphicx}

\begin{document}

\title{Understanding the Impact of openPMD on BIT1, a Particle-in-Cell Monte Carlo Code, through Instrumentation, Monitoring, and In-Situ Analysis}


\titlerunning{Understanding the Impact of openPMD on BIT1}
%
\author{Jeremy J. Williams\inst{1} \and Stefan Costea\inst{2} \and  Allen D. Malony\inst{3} \and  David Tskhakaya\inst{4}  \and Leon Kos \inst{2} \and Ales Podolnik\inst{4} \and  Jakub Hromadka \inst{4} \and  Kevin Huck\inst{3}  \and Erwin Laure\inst{5} \and Stefano Markidis\inst{1}}
\authorrunning{Jeremy J. Williams et al.}
%
\institute{KTH Royal Institute of Technology, Stockholm, Sweden \and LeCAD, University of Ljubljana, Ljubljana, Slovenia \and University of Oregon, Eugene, Oregon, The United States of America \and Institute of Plasma Physics of the CAS, Prague, Czech Republic \and
 Max Planck Computing and Data Facility, Garching and Greifswald, Germany  }
\maketitle              
\begin{abstract}
Particle-in-Cell Monte Carlo simulations on large-scale systems play a fundamental role in understanding the complexities of plasma dynamics in fusion devices. Efficient handling and analysis of vast datasets are essential for advancing these simulations. Previously, we addressed this challenge by integrating openPMD with BIT1, a Particle-in-Cell Monte Carlo code, streamlining data streaming and storage. This integration not only enhanced data management but also improved write throughput and storage efficiency. In this work, we delve deeper into the impact of BIT1 openPMD BP4 instrumentation, monitoring, and in-situ analysis. Utilizing cutting-edge profiling and monitoring tools such as gprof, CrayPat, Cray Apprentice2, IPM, and Darshan, we dissect BIT1's performance post-integration, shedding light on computation, communication, and I/O operations. Fine-grained instrumentation offers insights into BIT1's runtime behavior, while immediate monitoring aids in understanding system dynamics and resource utilization patterns, facilitating proactive performance optimization. Advanced visualization techniques further enrich our understanding, enabling the optimization of BIT1 simulation workflows aimed at controlling plasma-material interfaces with improved data analysis and visualization at every checkpoint without causing any interruption to the simulation.

\end{abstract}


\section{Introduction}
Particle-in-Cell (PIC) Monte Carlo (MC) simulations are critical for understanding plasma dynamics in fusion devices, requiring efficient data handling and analysis. Our prior work addressed the critical need for high-throughput parallel I/O in these simulations by integrating openPMD with the BIT1 code, enabling seamless streaming of particle and field information to storage systems. This integration not only improved data handling but also enhanced write throughput and storage efficiency. Building upon this, in this work, we investigate the impact of BIT1 openPMD BP4 instrumentation, monitoring, and in-situ analysis. We utilize state-of-the-art profiling tools like gprof, CrayPat, Cray Apprentice2, IPM, and Darshan to analyze BIT1's performance post-integration, uncovering insights into computation, communication, and I/O operations. Thorough instrumentation provides fine-grained insights into BIT1's runtime behavior, while immediate monitoring enhances understanding of system dynamics and resource utilization patterns, facilitating proactive performance tuning and optimization efforts. Advanced visualization techniques further aid in representing data flow, system interactions, and performance bottlenecks, empowering us to optimize BIT1 simulation workflows aimed at controlling plasma-material interfaces with improved data analysis and visualization at every checkpoint without causing any interruption to the simulation.

In this work, we focus on understanding the impact of openPMD enabling high-throughput parallel I/O in BIT1, achieved through comprehensive instrumentation, monitoring, and in-situ analysis. The contributions of this work include:

\begin{itemize}
    \item We identify the most computationally intensive parts by applying an I/O adaptor for the openPMD I/O interface that uses ADIOS2 BP4 as the I/O interface, which helps us understand the performance impact of running BIT1 openPMD BP4 on a single node.
    \item We apply profiling and monitoring techniques to evaluate the impact of using openPMD to implement parallel I/O compared to traditional file I/O in BIT1 when diagnostics are activated in strong scaling tests.
    \item We utilize a customized Python script with the openPMD API and ADIOS2 BP4 backend for real-time checkpoint access and visualization of BIT1 File I/O (from disk), without causing any interruption to the simulation.
\end{itemize}

\section{Background}
The PIC method serves as a numerical technique utilized in emulating plasma behaviors. This method governs particle dynamics across one to three dimensions in physical space and typically employs three dimensions in velocity space. In plasma edge scenarios, the PIC method often integrates MC routines to simulate particle collisions and their interactions with the walls of the plasma device chamber. The computational PIC cycle comprises five distinct phases: initiating plasma density calculations via particle-to-grid interpolation, executing a density smoothing operation to eliminate spurious frequencies, employing a field solver to tackle linear systems for electric and magnetic fields, managing particle collisions and wall interactions using MC techniques, and advancing particle positions and velocities over time, as detailed in~\cite{williams2023leveraging}.

BIT1 stands out as a tool of choice designed for accurately describing atomic processes and collisions during plasma-wall interactions. BIT1 is an electrostatic PIC code optimized for plasma edge modeling, with the key goal of enabling full-scale kinetic modeling of the plasma edge in next-generation fusion devices like ITER and DEMO. The input to BIT1 is a relatively small file (around 3 kB) that is read by all processes. The output analysis corresponds to two critical input parameters~\cite{williams2024parallelio}:

\begin{itemize}
    \item \textbf{mvflag}: Represents a flag for activating and enabling time-dependent diagnostics of plasma profiles and particle angular, velocity, and energy distribution functions. If > 0, it specifies the number of time steps over which the time-dependent diagnostics are averaged.
    \item \textbf{mvStep}: Counts the time steps for the interval between time-dependent diagnostics. 
\end{itemize}

While the original version of BIT1 boasts robust serial I/O functionality, the need for parallel I/O capabilities has become apparent, particularly as simulations scale up in size and complexity. BIT1's serial I/O encountered challenges beyond certain thresholds, becoming time-consuming and prone to file corruption, requiring the implementation of novel parallel I/O methods. Introducing new libraries and tools can enhance certain aspects of the code but may also introduce challenges in other areas. To address these issues and ensure the accuracy of output files while optimizing performance for extensive simulations, it's imperative to implement novel parallel I/O methods in BIT1. 

\subsection{ADIOS Version 2}
ADIOS2 (Adaptable Input/Output System 2) is a high-performance I/O library designed for managing data movement in scientific simulations and applications, offering flexibility and efficiency~\cite{godoy2020adios}. ADIOS2 has many support engines, BP4 (Binary Pack 4) stands as one of the supported data formats, optimized for performance and scalability, especially for large-scale runs. BP4 enables reduced storage requirements, faster read/write speeds, and compatibility with parallel I/O operations. When ADIOS2 is configured with the BP4 backend, it means that the library is tailored to utilize the BP4 format for storing and retrieving data, a configuration particularly beneficial in scientific computing scenarios where performance and scalability are paramount.


\subsection{openPMD Standard \& openPMD-api Integration}
The openPMD Standard, abbreviated as "open Particle-Mesh Data," presents a standardized format engineered for efficiently storing and exchanging data from scientific simulations involving particles and meshes~\cite{openPMDstandard}. Its streamlined format not only facilitates the storage and exchange of data but also ensures that vital metadata is retained, enabling effective interpretation and analysis~\cite{williams2024parallelio}. Moreover, openPMD supports a wide array of backends for data storage, including popular formats like HDF5, ADIOS1, ADIOS2, and JSON. This adaptability empowers scientists to seamlessly integrate openPMD into their existing workflows, choosing the backend that aligns best with their specific needs and preferences.



\section{Methodology \& Experimental Setup}
In this work, we aim to understand the impact of integrating openPMD with BIT1, determining its associated performance characteristics, we employ a suite of sophisticated profiling and monitoring tools. Specifically, we utilize the following tools:

\begin{itemize}
    \item \textbf{gprof} is an open-source profiling tool that collects execution time data and identifies the functions most frequently used by the processor. Since each MPI process generates a separate \textbf{gprof} output, these individual profiling results are consolidated into a single report encompassing all statistics.
    \item \textbf{CrayPat \& Cray Apprentice2} are powerful profiling tools for investigating and optimizing the performance of parallel applications on Cray architectures ~\cite{budiardja2018using}. \textbf{CrayPAT} is used to instrument the code, while \textbf{Cray Apprentice2} enables interactive, graphical performance analysis and visualization, specifically for our strong scaling experiments.
    \item \textbf{IPM} (Integrated Performance Monitoring) is a performance profiling tool that captures the computation and communication activities of parallel programs. It provides detailed reports on MPI calls and buffer sizes~\cite{fuerlinger2010effective}.
    \item \textbf{Darshan} is a performance monitoring tool specifically designed for analyzing serial and parallel I/O workloads~\cite{snyder2016modular}. We assess the I/O performance of BIT1 in terms of write throughput by using \textbf{Darshan} logs to extract data on high-throughput and the amount of data stored by each file on the file system.
\end{itemize}

\subsection{Use Case \& Experimental Environment}

We focus on exploring the impact of openPMD on enabling high-throughput parallel I/O in BIT1. Our simulations target neutral particle ionization resulting from interactions with electrons in upcoming magnetic confinement fusion devices such as ITER and DEMO. The scenario involves an unbounded unmagnetized plasma consisting of electrons, $D^+$ ions and $D$ neutrals. Due to ionization, neutral concentration decreases with time according to $\partial n / \partial t = n n_e R$, where $n$, $n_e$ and $R$ are neutral particles, plasma densities and ionization rate coefficient, respectively. We use a one-dimensional geometry with 100K cells, three plasma species ($e$ electrons, $D^+$ ions and $D$ neutrals), and 10M particles per cell per species. The total number of particles in the system is 30M. Unless differently specified, we simulate up to 200K time steps. An important point of this test is that it does not use the Field solver and smoother phases (shown in \cite{williams2023leveraging}).




We simulate and evaluate the impact of openPMD enabling high-throughput parallel I/O in BIT1 on the following three distinct systems:

\begin{itemize}
\item \textbf{Dardel}, an HPE Cray EX supercomputer, has 1270 compute nodes, each with two AMD EPYC™ Zen2 2.25 GHz 64-core processors, 256 GB DRAM. Nodes are connected via HPE Slingshot network (200 GiB/s) with Dragonfly topology. Storage includes a 12 PB Lustre File System (LFS) with 48 Object Storage Targets (OSTs). The OS is SUSE Linux Enterprise Server 15 SP3, with applications compiled using GCC 11.2, openPMD 0.15.2, ADIOS2 2.10.0 (Blosc and bzip2 enabled), and Cray MPICH 8.1.

\item \textbf{Discoverer}, a petascale EuroHPC supercomputer, has 1128 compute nodes, each with two AMD EPYC 7H12 64-Core processors, 256 GB DDR4 SDRAM (regular nodes), or 1TB DDR4 SDRAM (fat nodes). Nodes are connected via Ethernet Controller I350 (10 GiB/s) and Mellanox ConnectX-6 InfiniBand (200 GiB/s) with Dragonfly+ topology. Storage includes a 4.4 TB Network File System and a 2.1 PB LFS with 4 OSTs. The OS is Red Hat Enterprise Linux 8.4, with applications compiled using GCC 11.4.0 and MPICH 4.1.2.

\item \textbf{Vega}, a petascale EuroHPC supercomputer, has 960 compute nodes, each with two AMD EPYC 7H12 64-Core processors, 256 GB DDR4 SDRAM (80\% nodes), or 1TB DDR4 SDRAM (20\% nodes). Nodes are connected via Mellanox ConnectX-6 InfiniBand HDR100 (500 GiB/s) with Dragonfly+ topology. Storage includes a 23 PB Ceph File System (CephFS) and a 1 PB LFS with 80 OSTs. The OS is Red Hat Enterprise Linux 8, with applications compiled using GCC 12.3.0 and OpenMPI 4.1.2.1.
\end{itemize}


\subsection{BIT1 openPMD BP4 I/O Workflow}
As outlined by Williams et al.~\cite{williams2023leveraging,williams2024optimizing}, BIT1 performs serial I/O operations throughout each simulation. Similar to the process described in~\cite{poeschel2021transitioning}, a workflow has been established using specific ADIOS2 engines along with the requisite output extensions (\texttt{.bp}, \texttt{.bp4}, and \texttt{.bp5} respectively). For each extension, a distinct ADIOS2 file (or directory) is generated, containing one or multiple data files (\texttt{data.0, data.1 ... data.N, data.N+1}), a metadata file (\texttt{md.0}), an index table (\texttt{md.idx}), and, if enabled, a profiling file (\texttt{profiling.json}). BIT1 I/O workflow using openPMD with the ADIOS2 BP4 engine will employ the output extension directory, \texttt{data\_file.bp4}~\cite{williams2024parallelio}.


\section{Performance Results \& Analysis}
In this work, we investigate the impact of integrating openPMD with BIT1 using the ADIOS2 BP4 backend. 

\begin{figure}[!ht]
    \vspace{0cm} 
    \begin{center}
        \includegraphics[width=0.8\linewidth]{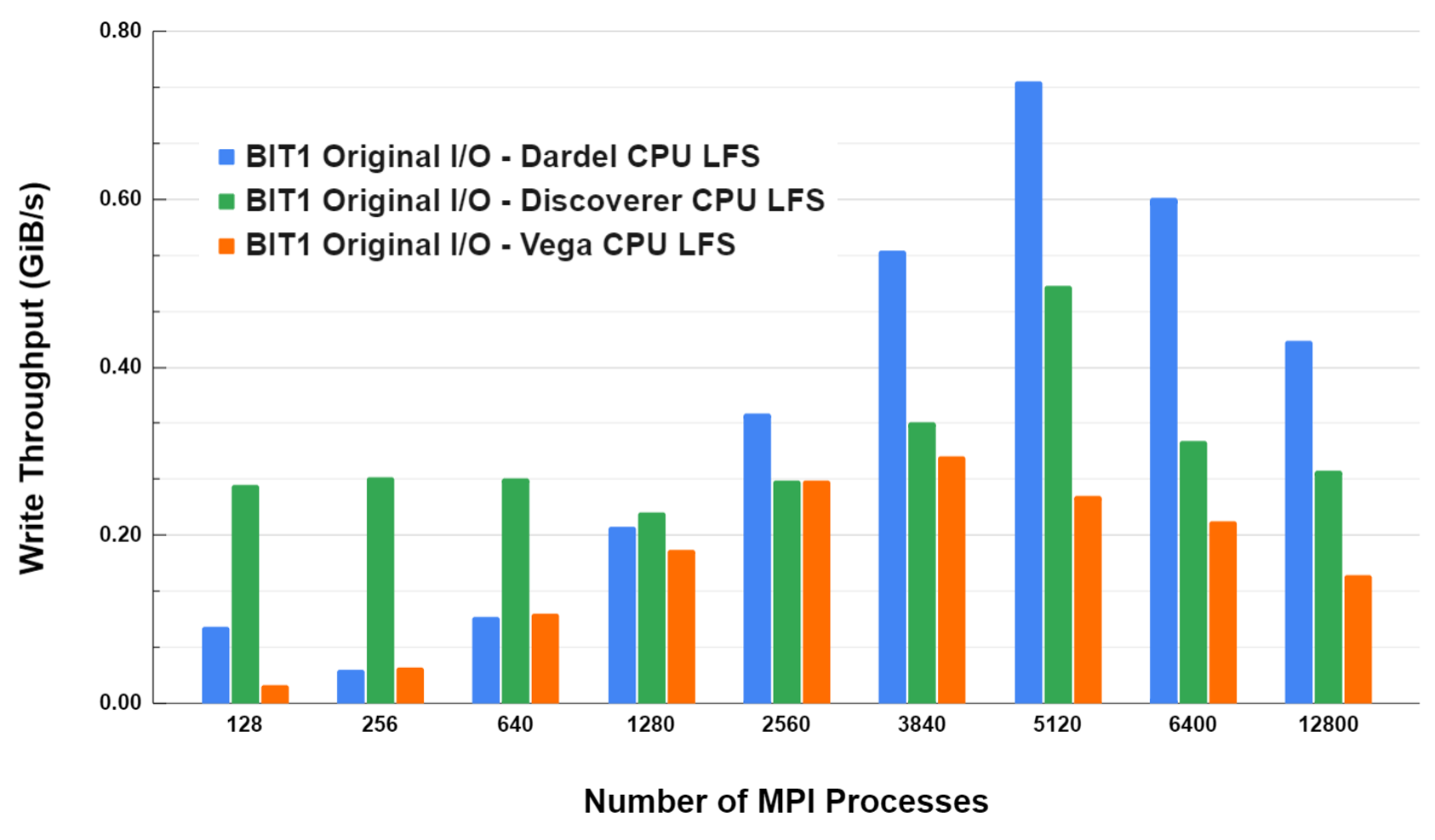}
        \caption{BIT1 Original File I/O Write Throughput, on Discoverer, Dardel and Vega CPU LFS, up to 100 Nodes (12,800 MPI processes), measured in GiB/s~\cite{williams2024parallelio}.} \label{darshan_BIT1_Original_IO}
    \end{center}
    \vspace{-0.8cm} 
\end{figure}

\subsection{BIT1 openPMD BP4 Instrumentation \& Monitoring} 

We begin by utilizing \texttt{Darshan}, a performance monitoring tool tailored for analyzing I/O workloads; we assess BIT1's write throughput by extracting data from \texttt{Darshan} logs. 

\begin{figure}[!ht]
    \vspace{0cm} 
    \begin{center}
        \includegraphics[width=0.8\linewidth]{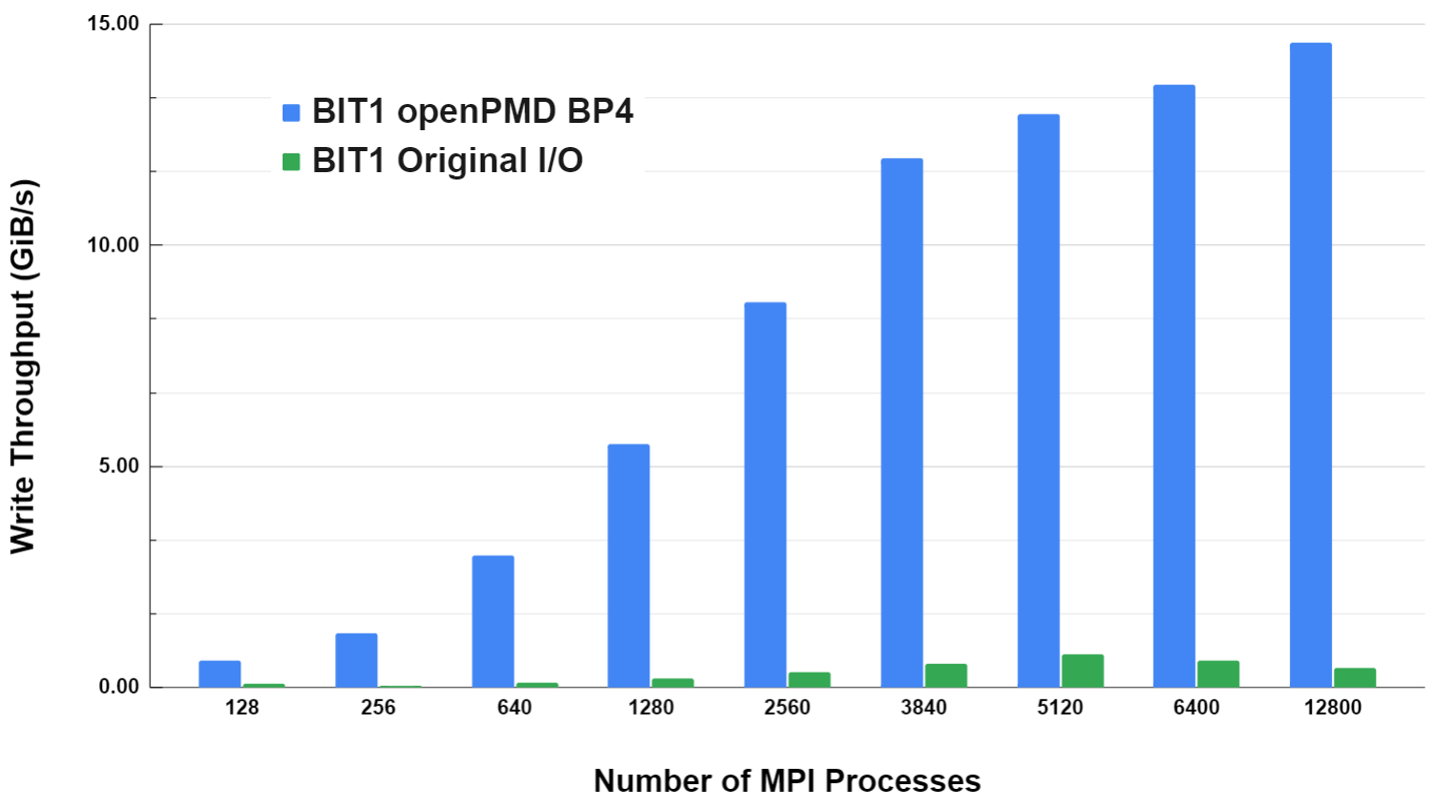}
        \caption{BIT1 openPMD BP4 and Original File I/O Write Throughput on Dardel up to 100 Nodes (12,800 MPI processes), measured in GiB/s~\cite{williams2024parallelio}.} \label{darshan_BIT1_IO_Before_and_After_Dardel}
    \end{center}
    \vspace{-0.8cm} 
\end{figure}

Fig.~\ref{darshan_BIT1_Original_IO} displays the Write Throughput (GiB/s) across three unique CPU LFS supercomputers: \texttt{Dardel}, \texttt{Discoverer}, and \texttt{Vega}. As the number of nodes increases, we observe varied performance trends across each system. \texttt{Discoverer's} performance shows fluctuations, with a slight initial increase followed by a decrease and then a minor increase again. \texttt{Dardel} exhibits generally increasing performance with the number of nodes. Notably, \texttt{Dardel} achieves the highest throughput among the three systems, reaching 0.74 GiB/s with 40 nodes. Vega's performance also demonstrates an upward trajectory overall, although with some fluctuations, especially evident at higher node counts. Despite these differences, \texttt{Dardel} consistently outperforms both Discoverer and Vega CPUs, making it the most promising option for further work. Its superior performance, as seen in Fig.~\ref{darshan_BIT1_IO_Before_and_After_Dardel} where BIT1 openPMD BP4 maintains stable throughput, indicates its suitability for tasks requiring high-throughput and efficiency. Therefore, it's recommended to continue our investigation on the \texttt{Dardel} CPU LFS Supercomputer to capitalize on its outstanding performance characteristics.

Next, we utilized \texttt{gprof}, an open-source profiling tool, to analyze execution time and identify the most frequently used functions across MPI processes. The consolidated \texttt{gprof} report offers a detailed performance analysis of BIT1 with and without openPMD.

\begin{figure}[!ht]
    \vspace{0cm} 
    \begin{center}
        \includegraphics[width=0.8\textwidth]{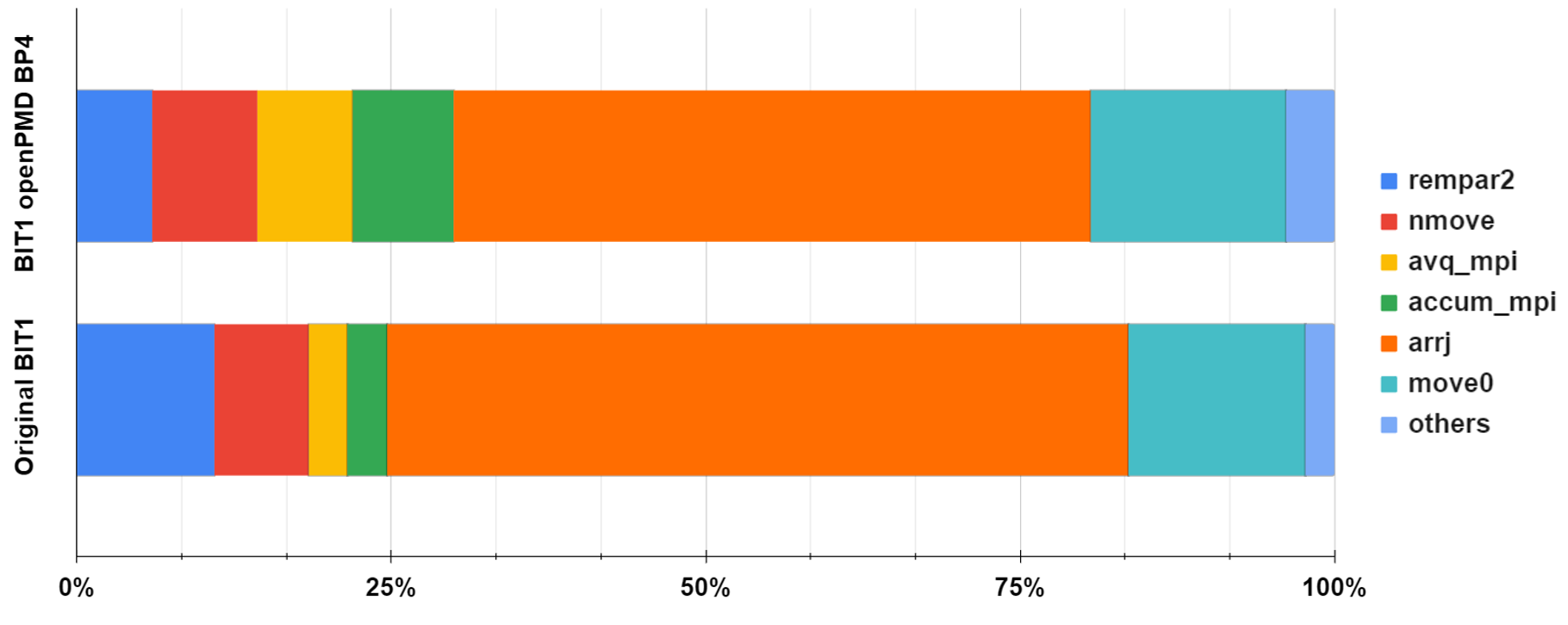}
        \caption{Percentage breakdown of the BIT1 Ionization functions where most of the execution time is spent using openPMD BP4 and Original File I/O~\cite{williams2023leveraging}. The \texttt{arrj} sorting function (in orange colour) is the function that takes most of the time, which have been reduced from 75\% to 65\%. The \texttt{gprof} tool have been used.} \label{function_breakdown}
     \end{center}
     \vspace{-0.8cm} 
\end{figure}

Fig.~\ref{function_breakdown} compares the performance of different operations between "BIT1 openPMD BP4" and "Original BIT1"  configurations, based on the percentage of total time spent on each operation. In the Original BIT1 configuration, the "arrj" function consumes 75.5\% of the total time. With BIT1 openPMD BP4, this drops to 65.5\%, indicating a 10\% improvement in efficiency, likely due to better data management and optimized processing.

Other notable observations include a slight increase in time spent on "move0" from 18\% to 20\%, possibly due to overhead from the new implementation. A significant decrease in "rempar2" from 14\% to 7.7\% suggests improved parallelization or data partitioning. There is a slight increase in "nmove" from 9.6\% to 10.8\%, which might be a trade-off for other improvements. Increases in "avq\_mpi" and "accum\_mpi" from 4\% to 9.8\% and 4\% to 10.4\%, respectively, indicate enhanced MPI operations due to better communication and data exchange mechanisms. Despite slight increases in some operations, the overall time distribution indicates a more balanced and optimized use of computational resources.

In addition to \texttt{gprof}, \texttt{CrayPat \& Cray Apprentice2} were employed to investigate the performance of the BIT1 openPMD BP4 application across various scales, ranging from small-scale runs on a single node to large-scale runs on up to 100 nodes. \texttt{CrayPat} facilitated code instrumentation, while \texttt{Cray Apprentice2} supported interactive, graphical performance analysis, and visualization, particularly for strong scaling experiments.

\begin{figure}[!ht]
    \vspace{0cm} 
    \begin{center}
        \includegraphics[width=\textwidth]{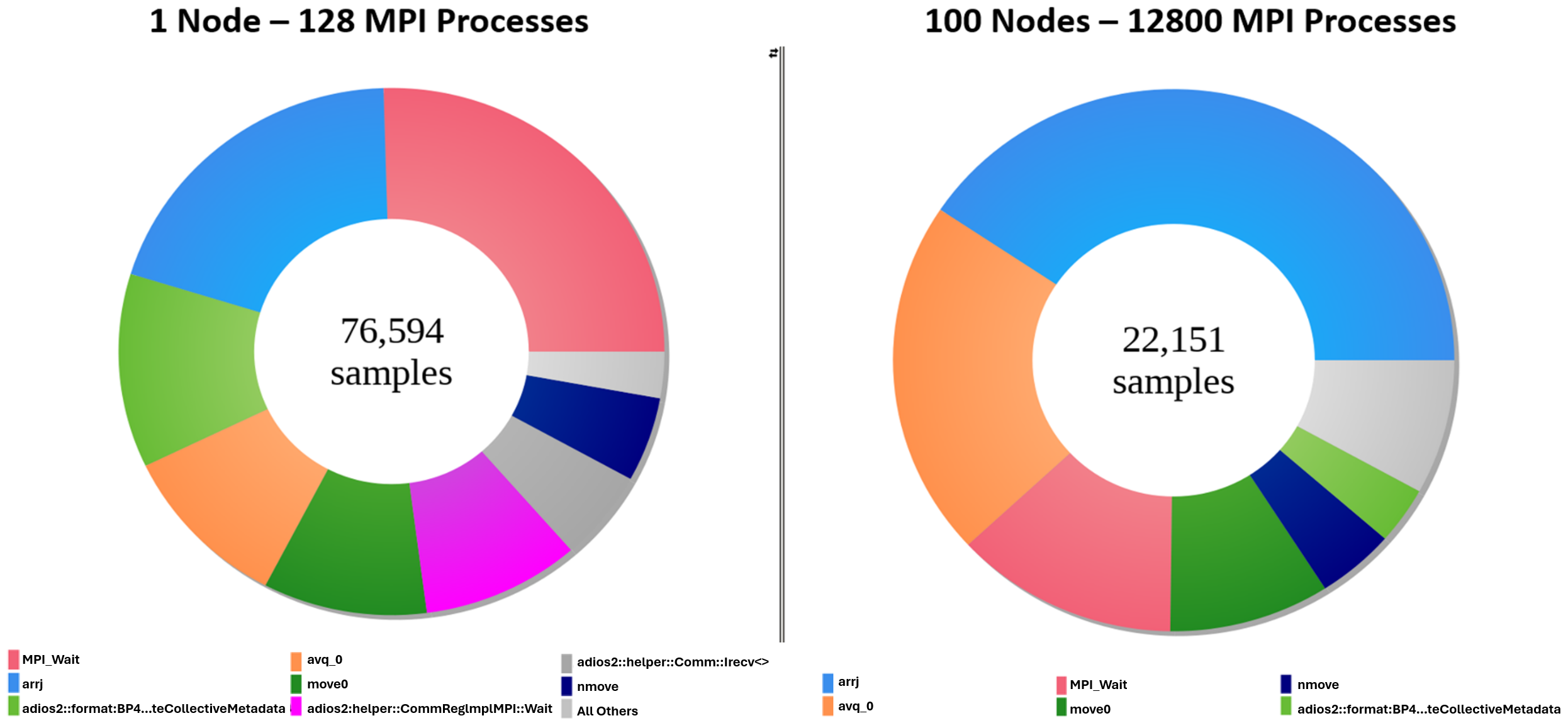}
        \caption{Percentage breakdown of the BIT1 openPMD BP4 Ionization call functions, where most of the execution time is attributed to these functions across small to large scale runs. As anticipated, the \texttt{arrj} sorting function (highlighted in blue) dominates the runtime, increasing from 19.8\% to 40\%. However, there is a notable decrease in the "MPI\_Wait" function, dropping from 25.5\% to 12.9\%. The analysis utilized the \texttt{CrayPat \& Cray Apprentice2} tools.} \label{cray_function_breakdown}
     \end{center}
     \vspace{-0.8cm} 
\end{figure}

Fig.~\ref{cray_function_breakdown} displays the performance of BIT1 on small to large scale runs, specifically focusing on the impact of the openPMD BP4 backend. For small-scale runs, \texttt{CrayPat} and \texttt{Cray Apprentice2} provided a detailed breakdown of function calls with significant exclusive sample hits, averaged across ranks. Notably, functions such as “arrj” (19.8\%), “adios2::AggregateCollectiveMetadata” (11.9\%), “MPI\_Wait” (25.5\%), “adios2::helper::CommReqImplMPI::Wait” (5.5\%), and “avq\_0' (10.2\%) exhibited substantial percentages of sample hits, revealing their impact on the overall performance of BIT1 openPMD BP4 runs. Additionally, “move0” (9.7\%) and “nmove” (5.2\%) were significant contributors to execution time. The remaining functions collectively accounted for 2.8\% of the sample hits.

Upon scaling up to 100 nodes, a different performance profile emerged. Despite the reduced total sample hits compared to small-scale runs, functions like “arrj” (40\%), “avq\_0” (21.3\%), “MPI\_Wait” (12.9\%), “move0” (9.3\%), and “nmove” (4.5\%) remained prominent contributors to the overall execution time. Notably, “adios2::AggregateCollectiveMetadata” (3.4\%) demonstrated a decrease in its impact. Interestingly, the “MPI\_Wait” function decreased from 25.5\% on a single node to 12.9\% on 100 nodes, which differs from the traditional expectation of increased MPI communication with node count. This decrease in MPI communication is attributed to the utilization of openPMD with the ADIOS2 BP4 backend and its aggregation capabilities, optimizing MPI communication, and enhancing overall performance compared to the original BIT1, as presented by Williams et al.~\cite{williams2023leveraging}.

\begin{figure} [!ht]
    \vspace{0cm} 
    \begin{center}
       \includegraphics[width=0.8\textwidth]{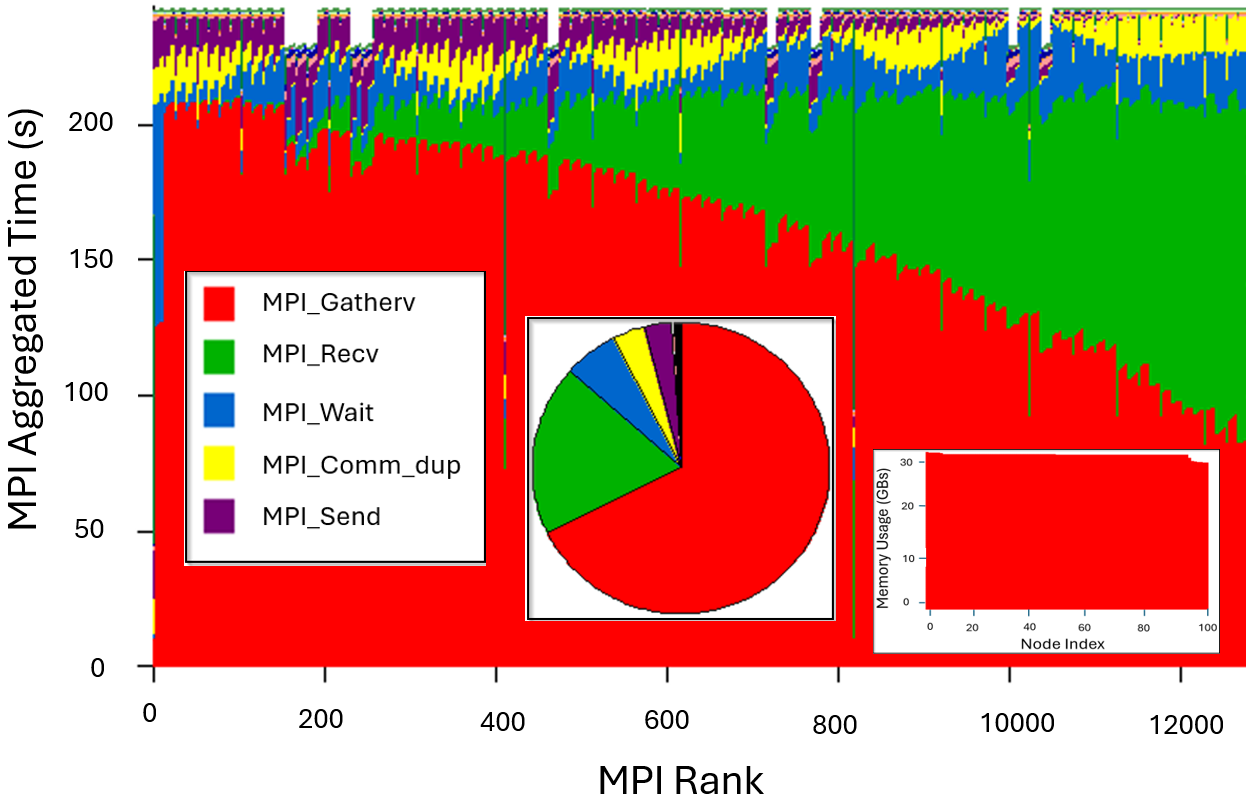}     
       \caption{MPI aggregated communication time for the BIT1 openPMD BP4 simulation on Dardel using 100 nodes, totaling 12,800 MPI processes.} \label{MPI_Communication_and_Memory_Usage}
    \end{center}
    \vspace{-0.8cm} 
\end{figure}

Based on the \texttt{CrayPat \& Cray Apprentice2} results, we further study overall MPI communication and load balancing in BIT1 openPMD BP4 simulations to investigate if there is a trade-off effect for this enhancement in the “MPI\_Wait” function. Fig.\ref{MPI_Communication_and_Memory_Usage} displays the MPI aggregated communication time for the BIT1 openPMD BP4 simulation on 100 nodes for a total of 12,800 cores. “MPI\_Gatherv” (67.65\%) dominates the communication time, indicating a need to optimize data gathering processes. Significant time is also spent in “MPI\_Recv” (19.04\%) and “MPI\_Wait” (5.76\%), indicating potential inefficiencies in message handling and synchronization on large runs. In Fig.\ref{MPI_Communication_and_Memory_Usage}, we also show the amount of memory consumed per node. There is a balanced use of compute nodes: the largest usage of memory per node is approximately 33~GB while the smallest is approximately 29~GB.  

\subsection{BIT1 openPMD BP4 In-Situ Analysis}
In-situ analysis facilitates real-time assessment of data directly within the environment where it is generated, without the need for extensive data transfers or storage. BIT1 can operate with minimal diagnostics, tracking the total particle number over time. Depending on input parameters, it can additionally log particle and power fluxes to the wall with minimal computational overhead. It also supports periodic system state saving for checkpointing and restoration.

\begin{figure}
    \vspace{0cm} 
    \begin{center}
        \includegraphics[width=\textwidth]{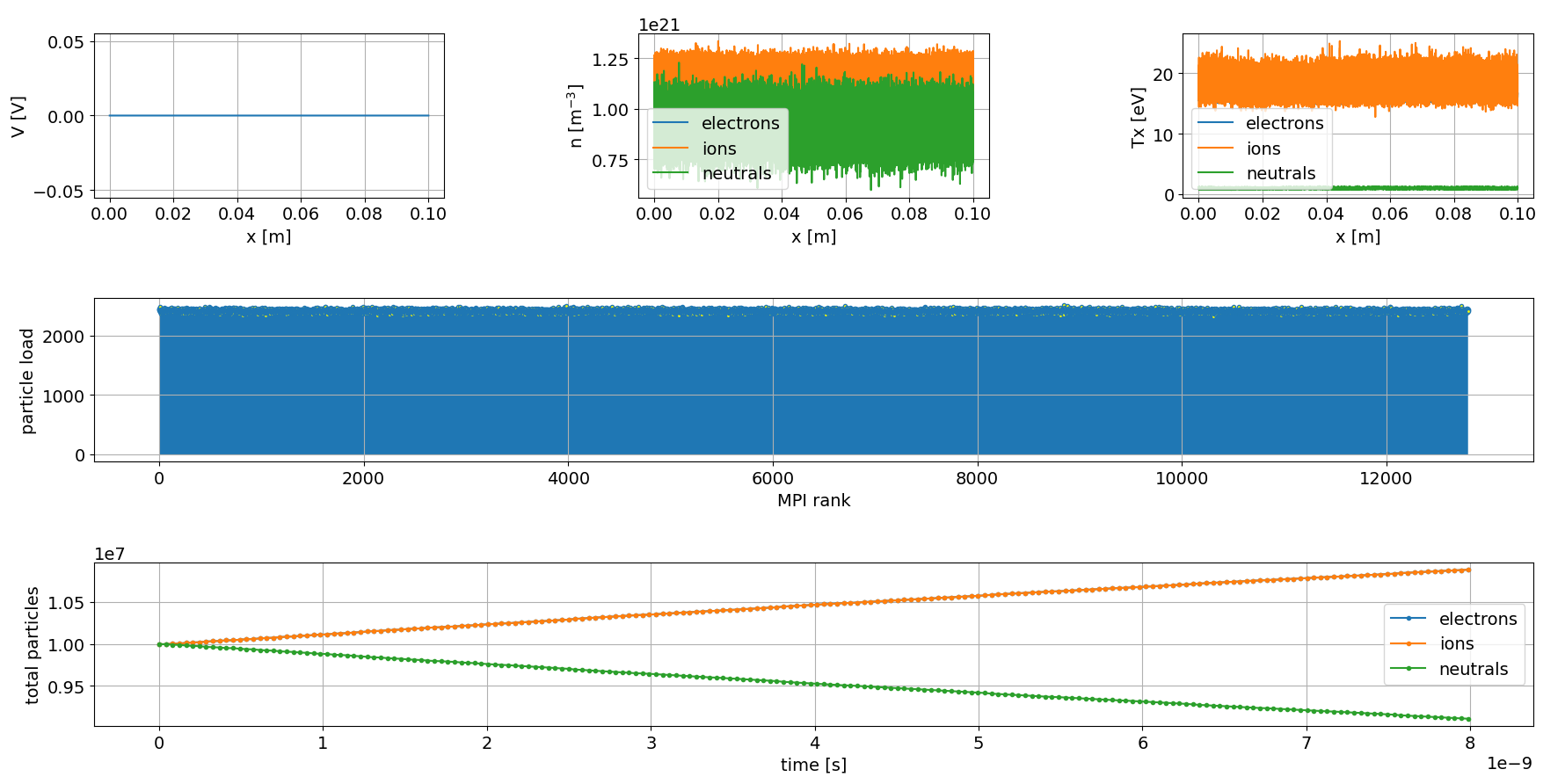}
        \caption{Performing a real-time checkpoint analysis on the Dardel CPU LFS for 12,800 MPI processes of the BIT1 openPMD BP4 simulation up to 200K time steps highlights a reduced set of plasma profiles, the particle load per MPI rank, and the time evolution of the total particles of each species.} \label{BI1_openPMD_BP4_visualization}
     \end{center} 
     \vspace{-0.8cm} 
\end{figure}

Fig~\ref{BI1_openPMD_BP4_visualization} shows a real-time checkpoint analysis of BIT1 File I/O stored in the output directory, \texttt{data\_file.bp4}, facilitated by a customized Python script utilizing the openPMD API and the ADIOS2 BP4 backend for 12,800 MPI Processes. Key visualizations include profiles of electric potential, plasma species densities, and temperatures, providing insights into plasma sheath presence and self-consistent electric field utilization.

Additionally, workload distribution among MPI ranks is visualized to ensure balanced computation load. Assessment of simulation steady-state is conducted through the time evolution of total particles of each species, aiding in understanding plasma sources and sinks dynamics. In the ionization case, the increase in electron and ion numbers due to the ionization of neutrals leads to a decrease in neutral particles over time. Using in-situ analysis offers immediate insights into system states, enabling efficient adjustments to simulation parameters. This real-time capability ensures the analysis occurs promptly at every checkpoint without causing any interruption to the simulation.

\section{Related Work}
PIC codes, crucial in simulating plasma physics and related fields, are increasingly undergoing performance and scalability enhancements. One approach involves integrating ADIOS2 and openPMD. ADIOS2, facilitating seamless data movement across various channels like files, networks, and direct memory, addresses the challenge of managing substantial data volumes generated by parallel simulations~\cite{godoy2020adios}. openPMD, an open-source initiative~\cite{openPMDstandard}, aims to standardize particle and mesh data file formats for diverse simulations, fostering interoperability and simplifying data analysis and visualization. Understanding openPMD's impact on BIT1, a PIC MC code, requires utilizing instrumentation, monitoring, and in-situ analysis techniques. Previous research highlights the significance of leveraging HPC profiling and tracing tools to analyze simulation performance, covering single-node, multiple-node, and I/O aspects~\cite{williams2023leveraging}. Additionally, optimizing BIT1 through OpenMP/OpenACC and GPU acceleration enhances its capabilities~\cite{williams2023leveraging}. Williams et al.~\cite{williams2023characterizing} demonstrated the importance of optimizing iPIC3D for large-scale 3D plasma simulations, offering practical recommendations to enhance performance and address Geospace Environmental Modeling magnetic reconnection challenges. Faj et al. ~\cite{faj2023mpi} analyzed Vlasiator's MPI performance, highlighting MPI nonblocking communication's dominance in communication time and advocating for OpenMP to eliminate intra-node communication, crucial for optimizing Vlasiator for Exascale machines. Poeschel et al. emphasized the openPMD-api as a valuable tool for describing scientific data according to the openPMD standard~\cite{poeschel2021transitioning,openPMDapi}. These studies collectively emphasize the importance of utilizing instrumentation, monitoring, and in-situ analysis capabilities to identify key areas for optimization, enhancement, and enablement in PIC MC codes like BIT1.

\section{Discussion and Conclusion}
The integration of openPMD with BIT1 marks a significant leap in plasma dynamics simulations for fusion devices. These advancements pave the way for further optimizations, boosting the accuracy and efficiency of models for plasma-material interfaces. Comprehensive instrumentation, monitoring, and visualization have provided valuable insights into BIT1's performance, particularly in computation, communication, and I/O operations.

This work emphasizes openPMD's crucial role in facilitating high-throughput parallel I/O in BIT1, crucial for handling vast data from plasma simulations. Profiling tools like \texttt{gprof}, \texttt{CrayPat}, \texttt{Cray Apprentice2}, \texttt{IPM}, and \texttt{Darshan} have helped identify areas for improvement. Comparing the original BIT1 setup with openPMD BP4 integration reveals significant efficiency gains, especially in data management and processing. The integration of the ADIOS2 BP4 backend with openPMD notably reduces write throughput bottlenecks and enhances performance across various computational platforms. Analyzing MPI communication and load balancing in BIT1 openPMD BP4 simulations offers insights into optimization strategies, particularly in data gathering and message handling. Visualization has played a pivotal role, providing intuitive representations of data flow and system interactions. In-situ analysis of electric potential profiles and workload distribution among MPI ranks has provided valuable real-time understanding of plasma dynamics and simulation behavior.

Future research will investigate the decrease in MPI communication when parallel I/O is considered. It can also explore integrating high-performance streaming using the Sustainable Staging Transport (SST) backend, extending real-time checkpoint analysis capabilities in memory rather than file I/O. Improvements in checkpoint restart and load balancing can enhance efficiency, resilience, and fault tolerance. Further integration can enable in-situ visualization with ParaView Catalyst 2 and ADIOS2 for efficient data transfer and non-blocking visualization capabilities, contributing to achieving exascale computing and addressing BIT1’s grand challenge of controlling plasma-material interfaces.


\vspace{2mm} 
\noindent \small{\textbf{Acknowledgments.} Funded by the European Union. This work has received funding from the European High Performance Computing Joint Undertaking (JU) and Sweden, Finland, Germany, Greece, France, Slovenia, Spain, and Czech Republic under grant agreement No 101093261. The computations/data handling were/was enabled by resources provided by the National Academic Infrastructure for Supercomputing in Sweden (NAISS), partially funded by the Swedish Research Council through grant agreement no. 2022-06725.



%
%
%

\end{document}